# Spin Polarization and Dispersion Effects in Emergence of Roaming Transition State for Nitrobenzene Isomerization[*]

Zhiyuan Zhang(张志远)[1,2], Wanrun Jiang(姜万润)[1,2], Bo Wang(王波)[1,2], Yanqiang Yang(杨延强)[3,†], and Zhigang Wang(王志刚)[1,2,†]

[1] *Institute of Atomic and Molecular Physics, Jilin University, Changchun 130012, China*

[2] *Jilin Provincial Key Laboratory of Applied Atomic and Molecular Spectroscopy (Jilin University), Changchun 130012, China*

[3] *Institute of Fluid Physics, China Academy of Engineering Physics, Mianyang 621900, China.*

Since roaming was found as a new but common reaction path of isomerization, many of its properties, especially those of roaming transition state ($TS_R$), have been studied recently on many systems. However, the mechanism of roaming is still not clear at the atomic level. In this work, we used first-principles calculations to illustrate the detailed structure of $TS_R$ in an internal isomerization process of nitrobenzene. The calculations distinctively show its nature of antiferromagnetic coupling between two roaming fragments. Moreover, the effect of dispersion is also revealed as an important issue for the stability of the $TS_R$. Our work provides a new insight from the view of electronic structure towards the $TS_R$ and contributes to the basic understanding of the roaming systems.

Roaming, as a newly discovered chemical reaction path, different from the conventional ones,[1] plays an important role in the isomerization process in many systems,[2-10] which has broadened our sight of view towards the reaction path. As a transition state (TS) in the roaming reaction path, roaming transition state ($TS_R$) take a critical place in the process,[11] for which its configuration and electronic structural properties have drawn a lot of attention. However, for experimental approaches, only indirect information could be observed from $TS_R$ and even normal TS. Besides, the differences between $TS_R$ and other conventional TS[12, 13] are not clearly revealed in the atomic level, for which we hope to understand the mechanism in this work.

Nitrobenzene is one of the typical energetic materials whose geometries and spectrum properties, isomerization and even dissociation channels have been studied theoretically and experimentally[14-19]. Recently, a new reaction channel in the photodissociation process of nitrobenzene was reported, suggesting a unimolecular isomerization process through a typical $TS_R$[20]. So far, most of the previous researches are focusing on the dynamic properties of the roaming systems, with certain atoms or functional groups roaming around the rest part and then dissociated or isomerized. However, more detailed mechanism of electronic structures in this roaming system still needs further investigations. Herein, our work carried out a theoretical investigation, with density functional theory (DFT), about the properties of roaming process of nitrobenzene, and particularly discovered the critical role that spin polarization interaction played in the $TS_R$, proving it to be the spin polarized roaming state (SPRS). Meanwhile, dispersion is also found to have a significant effect in stabilizing SPRS, obviously decreasing the distance between the functional groups. We hope that this result can be a valuable reference for a more regular recognition into the roaming mechanisms.

[*] Project supported by the Science Challenging Program (JCKY2016212A501) and National Natural Science Foundation of China (11374004), the High Performance Computing Center of Jilin University.
[†] Corresponding author. E-mail: wangzg@jlu.edu.cn(W. Z.) & yqyang@hit.edu.cn(Y. Y.)

In this work, based on the density function theory (DFT), the calculations are performed using B3LYP-D3 level[21] and 6-311+g(2df,2pd) basis set by means of Gaussian 09 package[22]. Two d and one f polarization functions are added to the C, N and O atoms, while two p and one d polarization functions are added to the H atom[23]. Diffuse functions are also introduced in the calculation[24]. The single point energy corrections are calculated under CCSD(T)/6-311++(2df, 2pd). Since the distance between the roaming separations is commonly larger than most of the bonding distance, dispersion, as an important issue in the intermolecular interaction, which were seldom discussed previously for the roaming systems, has become more effective for describing such systems. Therefore, the 3$^{rd}$ generation dispersion correction is introduced to B3LYP[25], namely B3LYP-D3, from which most of the conclusions are drawn. This result was also checked using M06-2X level[26]. The energies decomposition are calculated using ADF package[27].

Firstly, we obtained the roaming reaction path and especially the optimized $TS_R$ by means of the TS optimization and the intrinsic reaction coordinate (IRC) analysis under B3LYP level. The result shows that both $NO_2$ and $C_6H_5$ fragments are neutral with nearly one single spin election on each of them in the initial structure, showing a spin polarized singlet state structure. The distance between the ipso C atom in $C_6H_5$ fragment and the N atom in $NO_2$ ($D_{C-N}$) is 3.30 Å, corresponding to the 3.29 Å distance in the previous work[20].

Since dispersion is an important issue for describing intermolecular interactions, calculations with dispersion corrections under the same level (B3LYP-D3) are also performed. In the vibration calculation, one imaginary frequency was found in the spectrum for both cases, indicating that the stabilizations are qualitatively consistent no matter whether dispersion is introduced or not. However, after the dispersion correction is introduced in the calculation, the results show that $D_{C-N}$ were decreased by nearly 0.30 to 3.01 Å (2.96 Å for the M06-2X level), which has not been found in the conventional TS structure of the unimolecular isomerization. From this side, the special properties could also be seen.

In order to acquire the electronic structure of the roaming state, we carried out spin and charge density calculations, which are also important for understanding of electronic structure from wavefunction aspect. The result of spin density shows that, on each of the two fragments, the spin value is 0.98, corresponding to the two single electrons mentioned above. For the $C_6H_5$ fragment, most of the spin densities (91.0%) are localized on the ipso C atom, while for the $NO_2$ fragment, the spin density shows a more delocalized feature with 46.7% on the N atom and 26.6% on each of the O atoms (see Fig. 1(a)), which can be obtained similarly from M06-2X level as 89.5% for the ipso C atom in the $C_6H_5$ fragment, 48.4% and 25.8% on each N and O atoms, respectively. Meanwhile the electronic density difference analysis shows a rather "intra-fragment" feature, indicating a minor difference between the roaming structure and free $C_6H_5$ and $NO_2$ radicals, as shown in Fig. 1(b). Moreover, a general feature for the roaming reaction path could be concluded from these results that the single electrons are still localized on each fragment, in spite of that such an occupation pattern would cause unpaired electrons. This is an important property of eigenstate for roaming systems. Also, this is why only the spin unrestricted calculation could describe roaming reaction paths[11, 20].

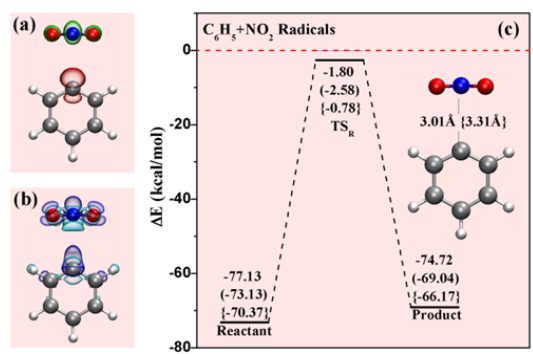

**Fig. 1.** The schematic potential surface (kcal/mol) of the intramolecular isomerization process through $TS_R$ and the electron structure of roaming nitrobenzene. (a) Spin density distributions of roaming-state nitrobenzene. The green and red isosurfaces suggest two different directions of spin density. (b) The electron density difference from two radical fragments $C_6H_5$ and $NO_2$ to the roaming nitrobenzene. The purple blue isosurface indicates the electron density increases, while the light blue isosurfaces means the decreases. (c) Roaming reaction path of nitrobenzene. The reactant is nitrobenzene, and the product is one of the isomers $C_6H_5NO_2$. The relative energies ($\Delta E$) compared with the corresponding free radicals form of $C_6H_5$ and $NO_2$ are marked. The numbers in the braces represent the results of B3LYP, those in the brackets indicate the results of B3LYP-D3, and those outside means the CCSD(T) results.

For further verification, the potential surface is obtained theoretically in the same level. As shown in Fig. 1(c), the energy barrier is 77.13 kcal/mol (73.13 kcal/mol for B3LYP-D3, 75.20 kcal/mol for M06-2X level) corresponding to the magnitude of similar roaming process reported previously[15]. Moreover, our results also indicate that the $TS_R$ is about 1.80 kcal/mol (2.58 kcal/mol for B3LYP-D3, 3.70 kcal/mol for M06-2X level) lower than the radical threshold, which is of similar magnitude to other roaming transition state systems[28]. On the other hand, for the fact without dispersion correction, the TS$_R$ turns out to be only 0.78 kcal/mol lower than the radicals. Therefore, it can be concluded that dispersion has a significant enhancement in the stability of such roaming systems.

Furthermore, in order to understand the properties of the interaction of the IRC reaction points, especially the TS$_R$, the energy decomposition analyses upon the configuration along the reaction path are carried out. According to the results, the bonding energy is -2.84 kcal/mol, among which the electronic absorption and Pauli repulsion are -4.80 kcal/mol and 5.70 kcal/mol respectively, indicating a 0.90 kcal/mol excluding effect for the steric case, while the orbital and dispersion are -1.71 and -2.03 kcal/mol respectively, as shown in Fig. 2. Therefore, dispersion takes 23.8% of the total attraction effects, which also proves its non-negligible contribution to the stability of such SPRS. For the other points on the IRC, which are all involved with spin polarizations, the results show that the total interaction energy of the TS$_R$ is the lowest, corresponding to its first order saddle point state. At the same time, the trends of the electronic interaction and the Pauli repulsion are similar, especially for those close to the TS$_R$, causing a lesser effect from steric interaction. This also indicates that the changing of the total interaction energy is mainly contributed by the orbital interaction.

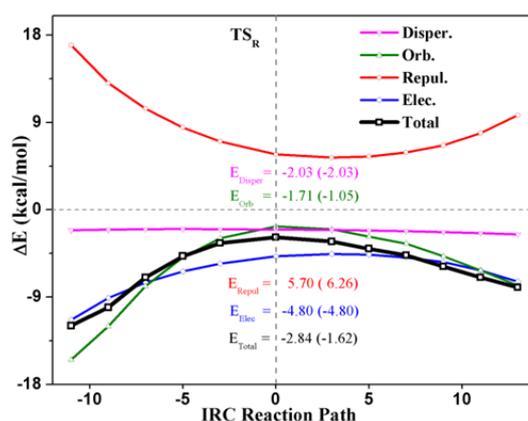

**Fig. 2.** The energy decompositions for part of the points along the IRC compared with those of the TS$_R$. The black line suggested the total bonding energy. The blue curve means the electronic absorption, and the red curve indicates the Pauli repulsion. The orbital interaction is represented by the green curve, and the dispersion is marked by the violet. The result of the energy decomposition for TS$_R$ with spin polarized singlet and triplet (in the bracket) states is listed in the corresponding color. Special technique was adapted in the calculation, introducing the restricted calculation as a "bridge", for the interaction from the restricted to the unrestricted fragments and compound can be easily obtained by specifically assigning the occupation[29]. Thus the result can be obtained with a simple subtraction as: $\Delta E = \Delta E_{com} - \Delta E_{frags}$, where $\Delta E_{com}$ and $\Delta E_{frags}$ both consist of electrostatic, repulsion, orbital and dispersion terms.

Moreover, considering that the triple state has the most similar electron structure to the spin polarized singlet state, we also compared the energy decomposition for these two cases. The result shows that the bonding energy for the spin polarized singlet is nearly twice as that of the triplet, which turns out to be -1.62 kcal/mol. Among those, the electrostatic term and the dispersion term are consistent, indicating the same spatial conformation. However, it should be noticed that from the ferromagnetic triplet to the antiferromagnetic spin polarized singlet, the Pauli repulsion is reduced by 0.56 from 6.26 kcal/mol, while the orbital interaction is increased by -0.66 from -1.05 kcal/mol. These could, on the other hand, prove that a spin polarized singlet structure for $TS_R$ involves a better stability.

Despite that the value of dispersion term is merely about 2 kcal/mol, dispersion interaction is still of great importance in stabilizing the conformation of $TS_R$, because the interaction distance in the roaming systems is far beyond the bonding area of functional groups. Moreover, the calculations above have indicated that dispersion corrected DFT would bring in about 0.30 Å decrease of $D_{C-N}$, which is a simple but effective example of its importance, and it is also a particular property compared with other transition states. Therefore, it could be concluded that the roaming is a special kind, which involves spin polarization, among the common reaction processes.

Since the spin-orbit coupling relativistic effects are important for open-shell systems, we further calculated the energy gaps of the highest occupied molecular orbital (HOMO) and the lowest unoccupied molecular orbital (LUMO) with both scalar and spin-orbit coupling effects. The results show that the gap of scalar effect is 67.24 kcal/mol (2.92 eV), which is in high consistence with that without relativistic effect. As expected, the gap of spin-orbit coupling approximation is 61.87 kcal/mol, relatively smaller than scalar cases. This is because that the frontier orbital energy levels are sensitive to the spin-orbit coupling, resulting in the LUMO and HOMO are respectively lowered and raised in energies. Therefore, the gap of HOMO and LUMO is decreased. The eigenvalue of $S^2$ operator $<S^2>$ is generally used to measure the spin contamination when compared with s(s+1), where s is 1/2 of the number of unpaired electrons, varying between systems with different spin multiplicities. For $TS_R$ of nitrobenzene, $<S^2>$ turns out to be 0.99, similar to the O(1D) + $H_2CO$ roaming system[30]. However, after the annihilation operation, $<S^2>$ was decreased to 0.08, which was in a reasonable range. Therefore, we believe that the unrestricted calculations are reliable.

Meanwhile, roaming transition states have been noticed and studied for quite a long time, whose energetic properties have been deeply discussed, while the detailed electronic properties at atomic level still need processing. Also, the DFT levels were expected to give a more unstable $TS_R$ compared with ab initial ones[11]. However, our calculation proves that the dispersion could significantly improve the stability of $TS_R$. It's noteworthy that, with dispersion correction introduced in the calculation, both the roaming distance and the energy of the $TS_R$ are obvious declined, indicating an more stable saddle point, which also suggest that dispersion does makes differences and should be considered in future roaming-related calculations. Nevertheless, although such SPRS configurations are saddle points that could be found in a certain distance, they are actually not a strict transition state compared to the conventional ones, due to their large amplitude and anharmonic motions. Therefore, they are more likely to be referred to as roaming regions.

In this work, we revealed the electronic structure of the SPRS for nitrobenzene. Both spin densities and electronic density difference proved that the roaming process of $TS_R$ involves with a spin polarized singlet structure. It should be noticed that apart from the instantaneous dissociating feature, the roaming transition state belongs to a bound reaction path which leads to the recombination of radical fragments and achieving the isomerization. The spin polarized electronic structure accounts for the bounded effect when two fragments are roaming. The energy decomposition also interpreted the stability of such an antiferromagnetic $TS_R$. Dispersion correction is proved to play an important role in the roaming processes by significantly decreasing the roaming distance. We hope that this work could provide a theoretical reference for the future investigations on the mechanism of roaming.

**Acknowledgment**
We would like to acknowledge the Science Challenging Program (JCKY2016212A501) and National Natural Science Foundation of China (11374004). Z. W. also acknowledges the High Performance Computing Center of Jilin University.